\documentclass[pra,twocolumn,showpacs,unsortedaddress]{revtex4-1}
\usepackage[dvips]{graphicx}
\usepackage{amsmath,amsfonts,amssymb}
\usepackage{dcolumn}
\usepackage{bm}
\usepackage{color}
\usepackage{epsfig}
\usepackage{hyperref}

\begin{document}

\def\jpb{J. Phys. B: At. Mol. Opt. Phys.~}
\def\pra{Phys. Rev. A~}
\def\prb{Phys. Rev. B~}
\def\prl{Phys. Rev. Lett.~}
\def\jmo{J. Mod. Opt.~}
\def\jetp{Sov. Phys. JETP~}
\def\etal{{\em et al.}}

\def\reff#1{(\ref{#1})}

\def\diff{\mathrm{d}}
\def\imagi{\mathrm{i}}

\def\halb{\frac{1}{2}}

\def\pabl#1#2{\frac{\partial #1}{\partial #2}}

\def\beq{\begin{equation}}
\def\eeq{\end{equation}}

\def\beqa{\begin{eqnarray}}
\def\eeqa{\end{eqnarray}}
\def\energy{{\cal E}}
\def\energykin{{\cal E}_\mathrm{kin}}

\def\eulere{\mathrm{e}}

\def\Ehat{\hat{E}}
\def\Ahat{\hat{A}}

\def\ket#1{\vert #1\rangle}
\def\bra#1{\langle#1\vert}
\def\braket#1#2{\langle #1 \vert #2 \rangle}

\def\vekt#1{\bm{#1}}
\def\vect#1{\vekt{#1}}
\def\vektr{\vekt{r}}

\def\makered#1{{\color{red} #1}}

\def\Im{\,\mathrm{Im}\,}

\def\varphic{\varphi_{\mathrm{c}}}

\def\Up{U_\mathrm{p}}
\def\vxc{v_\mathrm{xc}}
\def\vKS{v_\mathrm{KS}}
\def\vextop{\hat{v}_\mathrm{ext}}
\def\VC{V_\mathrm{c}}
\def\VHX{V_\mathrm{Hx}}
\def\VHXC{V_\mathrm{Hxc}}
\def\wop{\hat{w}}
\def\Gammaevenodd{\Gamma^\mathrm{even,odd}}
\def\Gammaeven{\Gamma^\mathrm{even}}
\def\Gammaodd{\Gamma^\mathrm{odd}}

\def\hamop{\hat{\cal H}}

\def\Vcalop{\hat{ V}}

\def\Tcalop{\hat{ T}}

\def\hamopKH{\hat{{\cal H}}_\mathrm{KH}}
\def\Hop{\hat{H}}

\def\Hintop{\hat{H}}

\def\Gop{\hat{G}}
\def\GKH{\hat{G}_\mathrm{KH}}
\def\Pop{\hat{P}}
\def\pop{\hat{p}}
\def\HopKS{\hat{H}_\mathrm{KS}}
\def\HKS{H_\mathrm{KS}}
\def\Top{\hat{T}}
\def\Tintop{\hat{T}_{\mathrm{int}}}
\def\Eop{\hat{E}}
\def\TopKS{\hat{T}_\mathrm{KS}}
\def\VopKS{\hat{V}_\mathrm{KS}}
\def\VKS{{V}_\mathrm{KS}}
\def\vKS{{v}_\mathrm{KS}}
\def\Ttildeop{\hat{\tilde{T}}}
\def\Ttilde{{\tilde{T}}}
\def\Vextop{\hat{V}_{\mathrm{ext}}}
\def\Vhxcop{\hat{V}_{\mathrm{hxc}}}
\def\Vext{V_{\mathrm{ext}}}

\def\Vintext{V_{\mathrm{ext}}^{\mathrm{int}}}

\def\Vopee{\hat{V}_{{ee}}}
\def\psiopdag{\hat{\psi}^{\dagger}}
\def\psiop{\hat{\psi}}
\def\vext{v_{\mathrm{ext}}}
\def\Vee{V_{ee}}
\def\vee{v_{ee}}
\def\nop{\hat{n}}
\def\Uop{\hat{U}}
\def\Vop{\hat{V}}
\def\Vintop{\hat{V}^{\mathrm{int}}}
\def\Wop{\hat{W}}
\def\bop{\hat{b}}
\def\bopdag{\hat{b}^{\dagger}}
\def\qop{\hat{q}}
\def\jop{\hat{j\,}}
\def\vHxc{v_{\mathrm{Hxc}}}
\def\vHx{v_{\mathrm{Hx}}}
\def\vH{v_{\mathrm{H}}}
\def\vc{v_{\mathrm{c}}}
\def\xop{\hat{x}}

\def\varphiexact{\varphi_{\mathrm{exact}}}

\def\fmathbox#1{\fbox{$\displaystyle #1$}}

\title{Periodicity of the time-dependent Kohn-Sham equation and the Floquet theorem}

\author{V.\ Kapoor}
\affiliation{Institut f\"ur Physik, Universit\"at Rostock, 18051 Rostock, Germany}

\author{M.\ Ruggenthaler}

\affiliation{Institut f\"ur Theoretische Physik, Universit\"at Innsbruck, Technikerstra{\ss}e 25, 6020 Innsbruck, Austria}

\author{D.\ Bauer}

\affiliation{Institut f\"ur Physik, Universit\"at Rostock, 18051 Rostock, Germany}

\date{\today}

\begin{abstract}
The Floquet theorem allows to reformulate periodic time-dependent problems such as the interaction of a many-body system with a laser field in terms of time-independent, field-dressed states, also known as Floquet states. If this was possible for density functional theory as well, one could reduce in such cases time-dependent density functional theory to a time-independent Floquet density functional theory. We analyze under which conditions the Floquet theorem is applicable in a density-functional framework. 
By employing numerical {\em ab initio} solutions of the interacting time-dependent Schr\"odinger equation with time-periodic external potentials we show that the exact effective potential in the corresponding Kohn-Sham equation is {\em not} unconditionally periodic. Whenever several Floquet states in the interacting system are involved in a physical process the corresponding Hartree-exchange-correlation potential is not periodic with the external frequency only. Using an analytically solvable example we demonstrate that, in general, the periodicity of the time-dependent Kohn-Sham Hamiltonian cannot be restored by choosing a different initial state. Only if the external periodic potential is sufficiently weak such that the initial state of the interacting system evolves adiabatically to a single, field-dressed state, the resulting Kohn-Sham system admits the application of the Floquet theorem.    
\end{abstract}
\maketitle

\section{Introduction}
The properties of multi-electron systems can in principle be predicted by solving the interacting many-body Schr\"odinger equation. However, numerical solutions are only feasible for small systems consisting of a few interacting electrons due to the exponential scaling of the computational demand with the number of particles. One possible way to overcome this so-called ``exponential wall'' \cite{kohnNobel} is density functional theory (DFT) \cite{DFT,DFTbooks}, which has been successfully applied to many-body systems in a wide range of areas in physics and chemistry. DFT is based on the existence of an energy functional whose mini\-miza\-tion yields the ground state density. This minimization is usually performed via the so-called Kohn-Sham (KS) construction \cite{KS} where the interacting multi-particle system is mapped to a unique system of non-interacting particles having the same ground-state density. The non-interacting problem decouples into (non-linear) one-particle equations with an effective Hamiltonian depending on the density. The advantage of DFT thus originates from the fact  that the solution of $N$ one-particle equations is less involved than solving one exponentially scaling $N$-particle problem. The crucial ingredient in the KS construction is the Hartree-exchange-correlation (Hxc) potential which, if it was exact, would comprise  all many-body effects. Its exact form is, however, unknown in general, and approximations have to be employed in practice.
 
Time-dependent density functional theory (TDDFT) extends DFT to time-dependent problems \cite{Runge,TDDFTbooks,Leeuwen}. The existence of a time-dependent KS system is, however, no longer based on a minimization principle but on the local force equation of quantum mechanics \cite{TDDFTbooks,ruggibauer}. The exact time-dependent Hxc potential has additional, subtle features: it depends on the initial state (both interacting and non-interacting)  and on the density at previous times (that is, it has ``memory'') \cite{inistatedep}. Therefore the construction of better approximations to the Hxc potential within TDDFT is much more involved than in DFT. This is even more of a problem as it turns out that the Hxc approximations known from DFT often fail when applied to TDDFT beyond linear response \cite{exampleswhereTDDFTfails}.

One might think that if TDDFT is employed to the study of multi-particle systems subject to time-periodic external potentials, e.g., an atom   interacting with a monochromatic laser field, one could involve the Floquet theo\-rem. Indeed,  in such situations the interacting many-body time-dependent Schr\"odinger equation is a partial differential equation with time-periodic coefficients and thus  admits a time-periodic basis. As a consequence, the problem can be converted into an infinite set of time-independent equations  by virtue of the Floquet theorem \cite{floquet,floquetclassics,floquetinbooks,floquetanalysis}.

Already at the beginning of the application of density-functional theory to time-dependent systems attempts were made to incorporate Floquet theory in a density-functional framework \cite{deb}. A minimization principle was proposed, which was perturbative in nature and hence valid only for weak and off-resonant fields. However, even if these conditions are met there are problems with defining a proper adiabatic limit, which is fundamental to the proposed minimization procedure \cite{Hone}. The problems arise due to the fact that Floquet theory maps the quasi-spectrum of the time-dependent problem into an interval of length $\omega$, i.e., the frequency of the periodicity employed. In any interval $I = \{x-\omega/2, x+\omega/2 \}$ for $x \in \mathbb{R}$ arbitrary we find infinitely many quasi-eigenenergies (they are dense in $I$) and thus infinitely many eigenfunctions around every point in the quasi-spectrum. A consequence of this is that there is no unique final state to which the system tends as the external perturbation is turned off adiabatically. In order to restore the adiabatic limit a truncation to a finite basis is usually employed, which is anyway unavoidable in practical calculations. 

In Refs.~\cite{TDDFTstates1,examplerev,floquetreview,TDDFTstates2} Floquet-DFT approaches were pursued for non-perturbative fields and later criticized in Refs.~\cite{floquetcritic,criticfinite} where the authors also suggested to embark upon the problem from a TDDFT point of view, thereby avoiding the minimization problem. The basic question then remains whether a Floquet basis can be found for the associated KS system, i.e., whether the KS Hamiltonian itself is periodic. Known explicit expressions for the exchange-correlation potential in the time-dependent KS Hamiltonian such as the adiabatic local density approximation or generalized gradient approximations \cite{DFTbooks,TDDFTbooks} have the feature that a periodic density will lead to a periodic KS Hamiltonian (with the same period) since the adiabatic Hxc potentials depend on the instantaneous density only. However, the density does not have to be  periodic and, in fact, it generally is not, as we will demonstrate in this work.  On the other hand, even if an approximate functional leads to an aperiodic KS potential because of, e.g., an aperiodic density,  this does not yet demonstrate the incompatibility of TDDFT and Floquet theory, because the unknown {\em exact} KS potential nevertheless could be periodic. In this work we will show by means of numerical and analytical counter examples that this, unfortunately, is not the case and thus TDDFT is, in general, not compatible with Floquet theory.

The paper is structured as follows. In Sec.~\ref{preliminaries} we review the basics of Floquet theory from a TDDFT perspective.  In Sec.~\ref{exactpotential} we compute the exact KS potential for a two-electron model system, present the Fourier-transformed exact KS potential, and investigate whether the Floquet theorem is applicable to the KS Hamiltonian. In Sec.~\ref{generalresults} an analytical example is given to analyze the initial-state dependence of the KS potential and its relation to the periodicity of the KS Hamiltonian. We conclude in Sec.~\ref{conclude}.

For simplicity, we restrict ourselves to one-dimensional systems in this work. Such systems are frequently used in the theory of laser-matter interaction because they can be solved numerically exactly, and they are known to capture many of the essential features of their three-dimensional analogs. All equations in this work can be straightforwardly extended to the  three-dimensional case.

Atomic units $\hbar=m_e=|e|=4\pi\varepsilon_0=1$ are used throughout unless stated otherwise.

\section{Basic Theory \label{preliminaries}}
Consider a system of $N$ interacting electrons governed by the Hamiltonian
\beq
\Hintop(t)=\Tcalop +\Vop_{\mathrm{ee}}+ \Vcalop(t) \label{inthamop}
\eeq
with, in position-space representation,  the kinetic energy operator
\beq \Tcalop=\sum_{i=1}^N -\frac{1}{2}\frac{\partial^2}{\partial x_i^2} \eeq
the interaction potential
\beq \Vop_{\mathrm{ee}}=\frac{1}{2}\sum_{i\neq j}^N v_{\mathrm{ee}}(|x_i-x_j|),\eeq
and the external potential 
\beq \Vcalop(t)=\sum_{i=1}^N v(x_i,t). \eeq 
We assume the interaction to be Coulombic. In one-dimensional models the Coulomb-interaction is usually smoothed by a softening parameter  $\epsilon>0$,  
\beq v_{\mathrm{ee}}(|x_i-x_j|)=\frac{1}{\sqrt{(x_i -x_j)^2+ \epsilon }}. \eeq
 We further specialize on external potentials consisting  of the interaction with a (static) nucleus of charge $Z$ and a laser field $E(t)$ in dipole approximation, i.e.,
\beq 
\label{externalfield}
v(x_i,t) = -\frac{Z}{{\sqrt{ x_{i}^2+\epsilon} }} +x_{i}E(t). \eeq
The eigenstates and eigenenergies of the laser field-free system at time $t=0$ are obtained via the solution of the time-independent Schr\"odinger equation
\beq \Hintop(t)\Psi(x_{1}\sigma_{1} \cdots x_{N}\sigma_{N})=\mathcal{E}\Psi(x_{1}\sigma_{1} \cdots x_{N}\sigma_{N}).\eeq
Here $\Psi(x_{1}\sigma_{1} \cdots x_{N}\sigma_{N})$ is an antisymmetric $N$-particle eigenfunction of the space and spin variables $x_i$, $\sigma_i$, and $\mathcal{E}$ is its eigenenergy. 
In order to obtain $\Psi(x_{1}\sigma_{1} \cdots x_{N}\sigma_{N},t)$ for $t>0$ one may solve the time-dependent Schr\"odinger equation (TDSE) 
\beq\imagi\partial_{t}\Psi(x_{1}\sigma_{1} \cdots x_{N}\sigma_{N}, t)=\Hintop(t)\Psi(x_{1}\sigma_{1} \cdots x_{N}\sigma_{N},t)\label{TDSE}
\eeq 
for a fixed initial state $\Psi_0(x_{1}\sigma_{1} \cdots x_{N}\sigma_{N})$. However, due to the ``exponential wall'' \cite{kohnNobel} it is computationally very challenging to solve this equation. In fact,  in the case of intense laser fields where the numerical grids need to be large it is feasible only for  $N\leq 3$.  

Now we turn our attention to the non-interacting KS system that, by construction, yields the same single-particle density $n(x,t)$ as the interacting system. For simplicity, we assume that we are dealing  with spin-neutral systems. The KS Hamiltonian then reads
\beq
\HopKS([n];t)=-\frac{1}{2} \frac{\partial^2}{\partial x^2} + v(x,t) + v_{\mathrm{Hxc}}([n];x,t),
\eeq
where $v(x,t)$ is the external potential \reff{externalfield} and $v_{\mathrm{Hxc}}([n];x,t)$ is the Hxc potential which is a functional of the single-particle density $n(x,t)$ (for notational simplicity we do not indicate the dependence on the initial states). The two potential terms combined are called the KS potential, i.e.,
\beq
v_{\mathrm{KS}}([n]; x,t)=v(x,t) + v_{\mathrm{Hxc}}([n];x,t).\label{KSpot}
\eeq

In what follows we assume that the external laser field is monochromatic with a period $\omega_1$.
The time-dependent KS equation reads
\beq \imagi\partial_{t}\Phi_{k}(x,t)=\HopKS([n];t)\Phi_{k}(x,t), \label{KS}\eeq
where $\Phi_{k}(x,t)$ is the $k$-th KS orbital for the KS particle with initial state $\Phi_{k}(x,0)$. 
The time-dependent one-particle density $n(x,t)$ then is
\beq
n(x,t)=\sum_{k=1}^{N} | \Phi_{k}(x,t)|^2 \label{density}.
\eeq

Now we make the basic assumption of any Floquet approach in a density-functional framework: if the Hamiltonian $\Hintop(t)$ describing the $N$ interacting electrons is periodic with the frequency $\omega_1$, i.e., $E(t+T)=E(t)$ with $T=2\pi/\omega_1$,  then we assume the same periodicity for the KS Hamiltonian as well. We neglect for the moment potential problems with respect to the non-linear nature of the KS equations, which will be discussed in detail in the subsequent Sections of this work. 

If the KS Hamiltonian is periodic with $T$ then, by virtue of the Floquet theorem, we can write the KS orbitals in a time-periodic (Floquet) basis $\{\phi_{\alpha}(x,t)\}_{\alpha \in \mathbb{N}}$ as
\beq \Phi_{k}(x,t)=\sum_{\alpha} c_{k \alpha} \eulere^{-\imagi\xi_{\alpha} t} \phi_{\alpha}(x,t), \label{psixt}\eeq
where the $\xi_{\alpha}$ are the so-called quasi-energies and $c_{k\alpha} = \braket{\phi_{\alpha}(t=0)}{\Phi_{k}(t=0)}$.
 Further, the $\phi_{\alpha}(x,t)$ are periodic in $T$, i.e., 
\beq \phi_{\alpha}(x,t)=\phi_{\alpha}(x,t+T). \label{periodicphi}\eeq
The Floquet orbitals $\phi_{\alpha}(x,t)$ fulfill the eigenvalue equation
\beq \hamop(t) \phi_{\alpha}(x,t) = \xi_\alpha\phi_{\alpha}(x,t) \label{fhami}\eeq
with 
\beq \hamop(t) = \HopKS([n];t) - \imagi\partial_t, \label{ffham}\eeq
i.e.,  $\xi_\alpha$ assumes the role of an eigenvalue and $\phi_{\alpha}(x,t)$ is  the corresponding eigenstate.
If so, also  
\beq \xi_\alpha'=\xi_\alpha+m\omega_1,\quad \phi_{\alpha}'(x,t)=\eulere^{\imagi m \omega_1 t} \phi_{\alpha}(x,t),\quad m\in\mathbb{Z} \eeq
are solutions of the eigenvalue equation \reff{fhami}.  Owing to the time periodicity of $\phi_{\alpha}(x,t)$ we can write
\beq \phi_{\alpha}(x,t)=\sum_{l} \varphi_{\alpha,l}(x) \eulere^{-\imagi l\omega_1 t}, \qquad l \in \mathbb{Z}.\label{fourexp}\eeq  
With Eqs.~\reff{psixt} and \reff{fourexp} the KS orbital can thus be written as,
 \beq \phi_{\alpha}(x,t)=\sum_{l\alpha} c_{k\alpha}\eulere^{-\imagi(\xi_{\alpha}+ l \omega_1) t}   \varphi_{\alpha,l}(x)   \label{fourex},\eeq
where the eigenstates $\{\varphi_{\alpha,l}(x)\}_{\alpha\in\mathbb{N},l\in\mathbb{Z}}$ form the time-independent Floquet basis.

 We divide the Hamiltonian $\HopKS([n];t)$ into a time-independent part 
\beq \hat{H}_0 = -\frac{1}{2}\frac{\partial^2}{\partial x^2} -\frac{Z}{{\sqrt{ x^2+\epsilon} }}, \eeq
the coupling to the monochromatic external field 
\beq x E(t)=v^{+}(x)\eulere^{\imagi \omega_1 t} +v^{-}(x)\eulere^{-\imagi \omega_1 t},\label{mono} \eeq
and $ v_{\mathrm{Hxc}}([n];x,t)$.
Since we tentatively assume time-periodicity of the whole KS Hamiltonian we can write
\begin{eqnarray}
 v_{\mathrm{Hxc}}([n];xt) =  \sum_{l}\eulere^{-\imagi l \omega_1 t}\left[v_{\mathrm{Hxc}}([n];x)\right]_l,\label{Hartree} 
\end{eqnarray}
$l \in \mathbb{Z}$.
Plugging the expansions \reff{fourexp}, \reff{mono}, and \reff{Hartree} in Eq.~\reff{fhami} we obtain the TDDFT-Floquet equations \cite{TDDFTstates1}
\begin{eqnarray} \lefteqn{(\xi_\alpha +l \omega_{1} -\Hop_0)\varphi_{\alpha,l}(x)} \label{flqeq} \\ & = & v^{+}(x)\varphi_{\alpha,l-1}(x) +   v^{-}(x)\varphi_{\alpha,l+1}(x)\nonumber \\ &+& \sum_{m}(v_{\mathrm{Hxc}}([n];x))_{l-m}(x)\varphi_{\alpha,l}(x). \nonumber  \end{eqnarray}

The index $l$ of a Floquet state $\varphi_{\alpha,l}(x)$ is known as the ``block index,'' which may be interpreted as the number of photons involved in the process under study (provided one arranges that the $l=0$-block adiabatically connects to the field-free situation). The Floquet equation \reff{flqeq}  couples any Floquet block $l$ to its neighboring blocks $l\pm 1$ via absorption or emission of a photon. Contributions of non-neighboring blocks may only be included through the Fourier-components of the Hxc potential. This is different from the  Floquet equations for the interacting TDSE which couple only neighboring blocks because $E(t)$ is the only time-dependent element in the TDSE Hamiltonian. However, in the TDSE case the Floquet basis functions depend on all spatial variables, not just on a single one as in the KS case.    

In principle,  Eq.~\reff{flqeq}  is an infinite-dimensional set of coupled partial differential  equations, in practice,  it is truncated  so that $l_{\min} \leq l \leq l_{\max}$ where $|l_{\min}|$ and $| l_{\max}|$ should be large enough to capture all the relevant processes in which photons are emitted or absorbed. 

If Eq.~\reff{flqeq} was valid, the periodic time-dependent many-body problem would be significantly simplified because the time-dependence had been eliminated via Floquet theory and the ``exponential wall'' via DFT.

\section{Periodic or aperiodic KS Hamiltonian? \label{exactpotential}}
In order to prove that Floquet theory is generally not applicable to TDDFT it certainly is sufficient to find one counterexample. However, a Floquet approach might still be useful as an approximative approach, especially given the fact that TDDFT in practice is itself  approximative anyway.    
Hence, we analyze under which circumstances the KS Hamiltonian is periodic or not. In order to do so we employ a widely used numerically exactly solvable   one-dimensional model Helium atom ~\cite{helium,exactpotential2,rabidieter}. In this model both electrons move along the laser-polarization direction only, and the Coulomb interaction is replaced by a soft-core potential as introduced in Sec.~\ref{preliminaries}. The TDSE Hamiltonian of the model system thus corresponds to the Hamiltonian \reff{inthamop} with $N=2$. The smoothing parameter was $\epsilon=1$, as, e.g.,  in \cite{rabidieter}.

The initial TDSE state is chosen to be the spin-singlet ground state of the interacting system
\beq
\Psi_0(x_1 \sigma_1, x_2 \sigma_2) = \Psi_0(x_1,x_2) \frac{1}{\sqrt{2}} \left(\ket{\uparrow_1} \ket{\downarrow_2} -\ket{\downarrow_1} \ket{\uparrow_2}  \right).
\eeq
Since the Hamiltonian is spin-independent, the system remains also during the dipole interaction with a laser field in a spin-singlet configuration, and we can concentrate on the symmetric spatial part  $\Psi_0(x_1,x_2)$ of the wave function only. The TDSE~\reff{TDSE} is solved numerically using the Crank-Nicolson propagator to obtain the time-dependent spatial wavefunction $\Psi(x_1,x_2,t)$. 

In Ref.~\cite{floquetanalysis} we introduced a method to extract the populated Floquet states of the interacting system directly from  $\Psi(x_1,x_2,t)$. 
By controlling the laser parameters we can either have an adiabatic evolution of the field-free state $\Psi_0(x_1,x_2)$ to a field-dressed (Floquet) state or a non-adiabatic one, where several Floquet states are populated. The laser intensity, frequency and the ramping time decide on the adiabaticity of the time-evolution of the interacting system. For adiabatic evolution we have in the TDSE-Floquet calculation only one relevant Floquet-state index $\alpha$ in the TDSE analog of \reff{fourex},
\beq \Psi(x_1,x_2,t)=\sum_{\alpha} c_{\alpha}\eulere^{-\imagi\xi_{\alpha} t} \sum_l \eulere^{-\imagi l\omega_1 t} \varphi_{\alpha,l}(x_1,x_2)   \label{fourexTDSE}.\eeq
Hence, in this case 
\beq \Psi(x_1,x_2,t) \sim \eulere^{-\imagi\xi_{\alpha} t} \sum_{l}   \eulere^{-\imagi l\omega_1 t} \varphi_{\alpha,l}(x_1,x_2), \eeq
and the density $n(x,t)=2\int\diff x'\, |\Psi(x,x',t)|^2$ will only have frequency components  proportional to multiples of the laser frequency $\omega_1$.  The KS Hamiltonian depends on the density. If the KS potential is periodic with respect to integer multiples of the laser frequency there would be no problem because $\HopKS([n(t+T)];t+T)=\HopKS([n(t)];t)$, and thus the Floquet theorem still holds.  Instead, fractional harmonics or,  even worse,  incommensurate frequencies in $\HopKS([n(t)];t)$ would render the Floquet theorem inapplicable. If more than one Floquet state is populated, say $\alpha=\alpha_1$ and $\alpha_2$, the Fourier-transformed density $n(x,\omega)$ will  also have frequency components  proportional to the inverse of the quasi energy difference $|\xi_{\alpha_2} - \xi_{\alpha_1}|$. It would be mind-boggling if the unknown {\em exact} $v_\mathrm{xc}([n];x)$ was able to remove such frequencies from  $\HopKS([n(t)];t)$. However, in order to {\em prove} that in general the exact $v_\mathrm{xc}([n];x)$ contains frequency components different from $\omega_1$ we construct the exact $v_\mathrm{xc}([n];x)$  explicitly in the following for both the adiabatic as well as the non-adiabatic evolution of the field-free state to the field-dressed states.

Once we have obtained $\Psi(x_1,x_2, t)$ by solving the TDSE \reff{TDSE}  we can construct the exact KS  orbital and the potential following Refs.~\cite{exactpotential2,exactpotential1}. In the two-electron spin-singlet case the KS wave function consists of only one spatial orbital $\Phi(x,t)$, i.e.,
\begin{align*}
\Phi(x_1 \sigma_1 &,x_2 \sigma_2,t) 
\\
&= \Phi(x_1,t)\Phi(x_2,t) \frac{1}{\sqrt{2}} \left(\ket{\uparrow_1} \ket{\downarrow_2} -\ket{\downarrow_1} \ket{\uparrow_2}  \right).
\end{align*}
The KS orbital can be written as
\beq \Phi(x,t)=\sqrt{n(x,t)/2}\ \eulere^{\imagi S(x,t)}, \label{KSdensphase} \eeq
where $n(x,t)$ is the exact particle density and $S(x,t)$ is the exact phase of the KS orbital. The expression for the phase in terms of density is given by the continuity equation as \cite{exactpotential1, QR}
\begin{align}
\label{continuity}
- \partial_x \left[ n(x,t) \partial_x S(x, t)\right] = \partial_t n(x,t).
\end{align}
Equation \reff{KS} can be inverted to write the KS potential in terms of the KS orbital as \cite{exactpotential1}
\begin{align}
\label{potentialvarphi}
v_{\mathrm{KS}}&(x,t) = \frac{\imagi \partial_t \Phi(x,t)+ \frac{1}{2} \partial_{x}^2\Phi(x,t)}{\Phi(x,t)}  \nonumber
\\
&=\frac{1}{2} \frac{\partial_x^2 \sqrt{n(x,t)}}{\sqrt{n(x,t)}} -\partial_t S(x,t) - \frac{1}{2}\left[  \partial_x S(x,t)\right]^2.
\end{align}
The imaginary part of the potential is zero due to the continuity equation~(\ref{continuity}). The density $n(x,t)$ and the phase $S(x,t)$ are computed from $\Psi(x_1,x_2,t)$ \cite{exactpotential2}, and by the above construction we obtain the exact KS potential. Such a straightforward construction is possible only if we have a single spatial orbital. In the general case of several KS orbitals one would need to employ a computationally more demanding fixed-point method, as demonstrated in Refs.~\cite{Godby, Soeren}. Once the exact KS potential is computed, it is Fourier transformed in time to investigate its periodicity.

Besides the basic problem of the periodicity of the KS potential for a given interacting density, there is the inherent non-linearity of the KS scheme. Even though the exact KS potential might be periodic for a certain problem, it is far from obvious that one can employ a Floquet-based KS scheme to predict it. For instance, although an adiabatic approximation, e.g., in the two-electron spin-singlet case the exact exchange-only approximation $v_\mathrm{Hx}^\mathrm{(exact)}([n];x)=\int\diff x'\, [n(x',t)/2]/\sqrt{(x-x')^2+\epsilon}$, does inherit the periodicity of the density, it is not guaranteed that the non-linear KS equations produce a periodic $n(x,t)$. This becomes obvious when we consider the iterative solution of the KS equations, where we start with an initial guess for the density that is periodic with $\omega_1$. We then have a periodic KS Hamiltonian from which we can (since in every iterative step we have a linear partial differential equation) infer a Floquet basis. We then solve the resulting linear equations and obtain a new density. This density will in general not be periodic and we no longer find a Floquet basis with period $\omega_1$ only. This makes the problem of the non-linearity in connection with a Floquet approach evident.
   
\subsection{Adiabatic and periodic example}
\begin{figure}\includegraphics[width=0.5\textwidth]{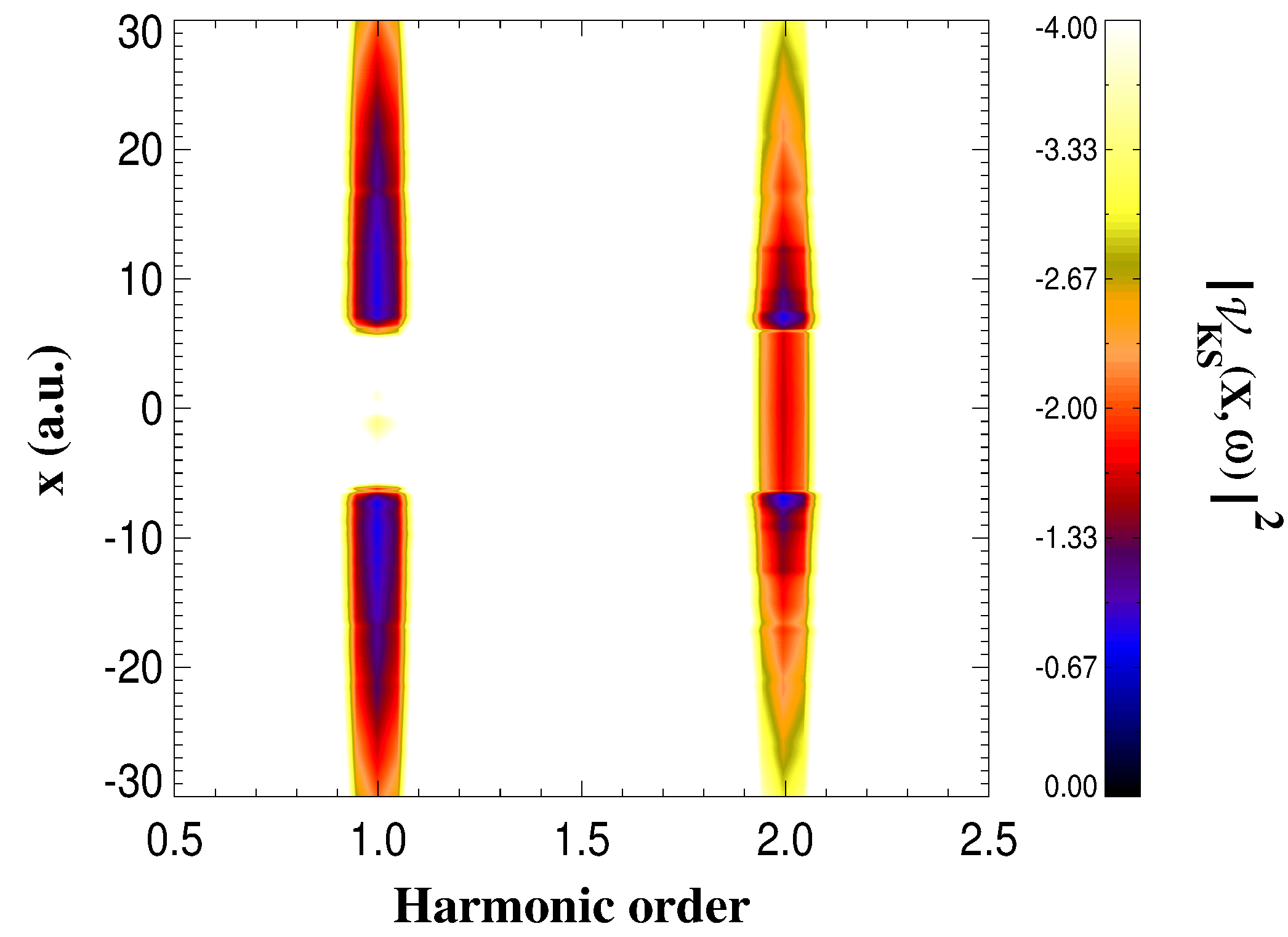} 
\caption{(Color online) Logarithmically scaled plot of $\arrowvert\vKS(x,\omega)\arrowvert^2$ for $\omega_1=0.056$, $\hat{E}=0.063$, two-cycle ramp-up, and $16$ cycles constant amplitude. Notice that only harmonics of the laser frequency are present over a dynamic range of four orders of magnitude. Superpositions of Floquet states do not play a role, the dynamics are sufficiently adiabatic, the Floquet theorem is applicable to $\HopKS([n(t)];t)$.  \label{fig:800nm}}
\end{figure}

First we consider an $800$-nm ($\omega_1=0.056$) laser pulse with two cycles ramp-up and $16$ cycles of constant amplitude. The electric field amplitude is $\hat{E}=0.063$, corresponding to a laser intensity of $1.4\times 10^{14}$~W/cm$^2$. It turns out that in this case the density dynamics are periodic with the laser period. In Fig.~\ref{fig:800nm} we plot the exact $ \arrowvert \vKS(x,\omega)\arrowvert^2 $ over four orders of magnitude vs the harmonic order $\omega/\omega_1$. Only harmonics of the laser frequency at all space points of the KS potential are visible. The Floquet theorem is applicable in this case, as  $\HopKS([n(t+T)];t+T)=\HopKS([n(t)];t)$ to a high degree of accuracy.

\subsection{Non-adiabatic and aperiodic example}\label{nonadiabaticexample}
\begin{figure}
\includegraphics[width=0.5\textwidth]{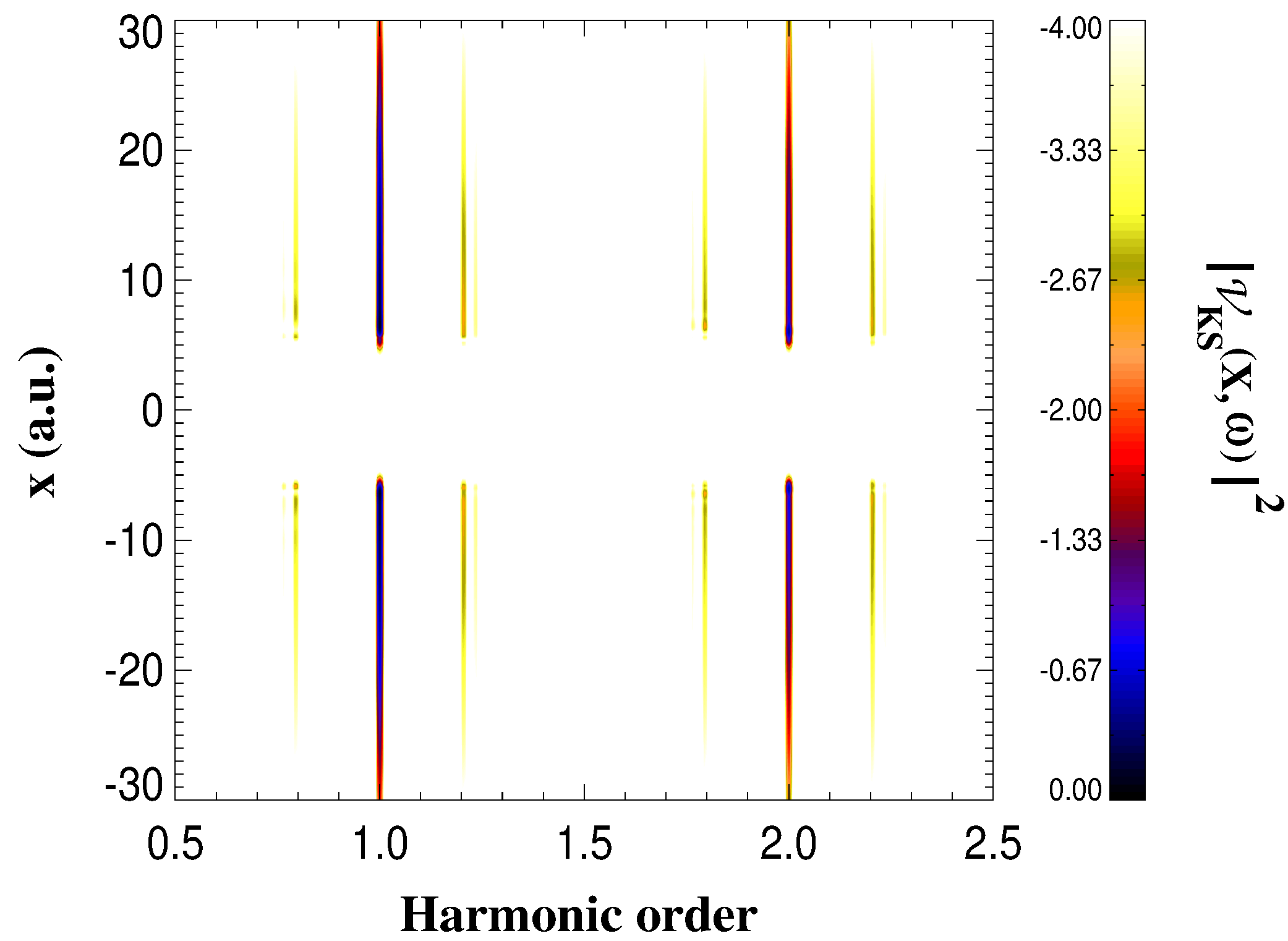} 
\caption{(Color online) As Fig.~\ref{fig:800nm} but for $\omega_1=2.6$,  $\hat{E}=0.34$, four-cycle ramp-up, and $172$ cycles constant amplitude. The Fourier-transformed potential displays anharmonic frequency components, i.e., it is aperiodic.\label{fig:3nm}}
\end{figure}

As a second example  we chose a short-wavelength $17.5$-nm  ($\omega_1=2.6$) laser pulse with four cycles ramp-up and $172$ cycles of constant amplitude. The electric field amplitude $\hat{E}=0.34$ corresponds to a laser intensity of $4 \times 10^{15}$~W/cm$^2$. The fast ramping induces a non-adiabatic time-evolution and results in a superposition of Floquet states in the TDSE result. The exact KS potential oscillates with periods related to the inverse of the quasi energy differences.  In Fig.~\ref{fig:3nm}, this new timescale manifests itself as side bands around the multiples of the laser frequency. The  quasi energy differences are determined by the field-free spectrum of the system under study and by the ac Stark shifts so that it may well happen that they are irrational fractions or multiples of $\omega_1$. In that case even a $T' > T=2\pi/\omega_1$  for which   $\HopKS(t+T')=\HopKS(t)$ does not exist.

\subsection{Resonant interaction}\label{resonantinteraction}
 When the laser is tuned to the exact resonance between the initial (ground) state and a dipole-accessible excited state, Rabi-oscillations set in, typically on a time scale that is much longer than the laser period so that for the Rabi frequency $\Omega$ one has $\Omega\ll\omega_1$. In this case the density is periodic with the Rabi-frequency $\Omega$, not with the laser frequency $\omega_1$. At time $T_{1/2}=\pi/\Omega$ the upper state is populated, at time $2\pi/\Omega$ the initial state is populated again. The Rabi-frequency depends on the electric field amplitude $\hat{E}$  of the laser and the transition dipole matrix element $\mu_{01}$ between the two bound states involved. Rabi-oscillations are not captured in TDDFT when known and practicable adiabatic exchange-correlation potentials are used. Of course, the density dynamics between the two states are correctly described when the exact KS potential is used, for instance for the numerically exactly solvable model-He system employed in this work. It is known that after the time $T_{1/2}$, when the single-particle density $n(x,T_{1/2})$ is that of the {\em excited interacting system}, the exact KS potential is the {\em ground state potential} to that density  \cite{rabidieter,helbig}. In fact, there is no stationary state in the KS potential to which the population may be transferred. Hence, the exact KS system governs the dynamics by an ``adiabatic deformation'' of the ground state density. Despite this extremely simple ``Rabi-flopping'' dynamics, resonant interactions are among the worst cases for TDDFT with known and practicable exchange-correlation potentials. 

It is well known that a Floquet treatment of the TDSE leads to avoided crossings of the two field-dressed state energies when plotted as a function of laser frequency \cite{floquetinbooks}. At exact resonance the two Floquet states are equally populated and separated in energy by $\hbar \Omega$. Hence, resonant interaction is a prime example where a superposition of Floquet states plays a role even if the laser pulse was turned on adiabatically.

\begin{figure}
\includegraphics[width=0.5\textwidth]{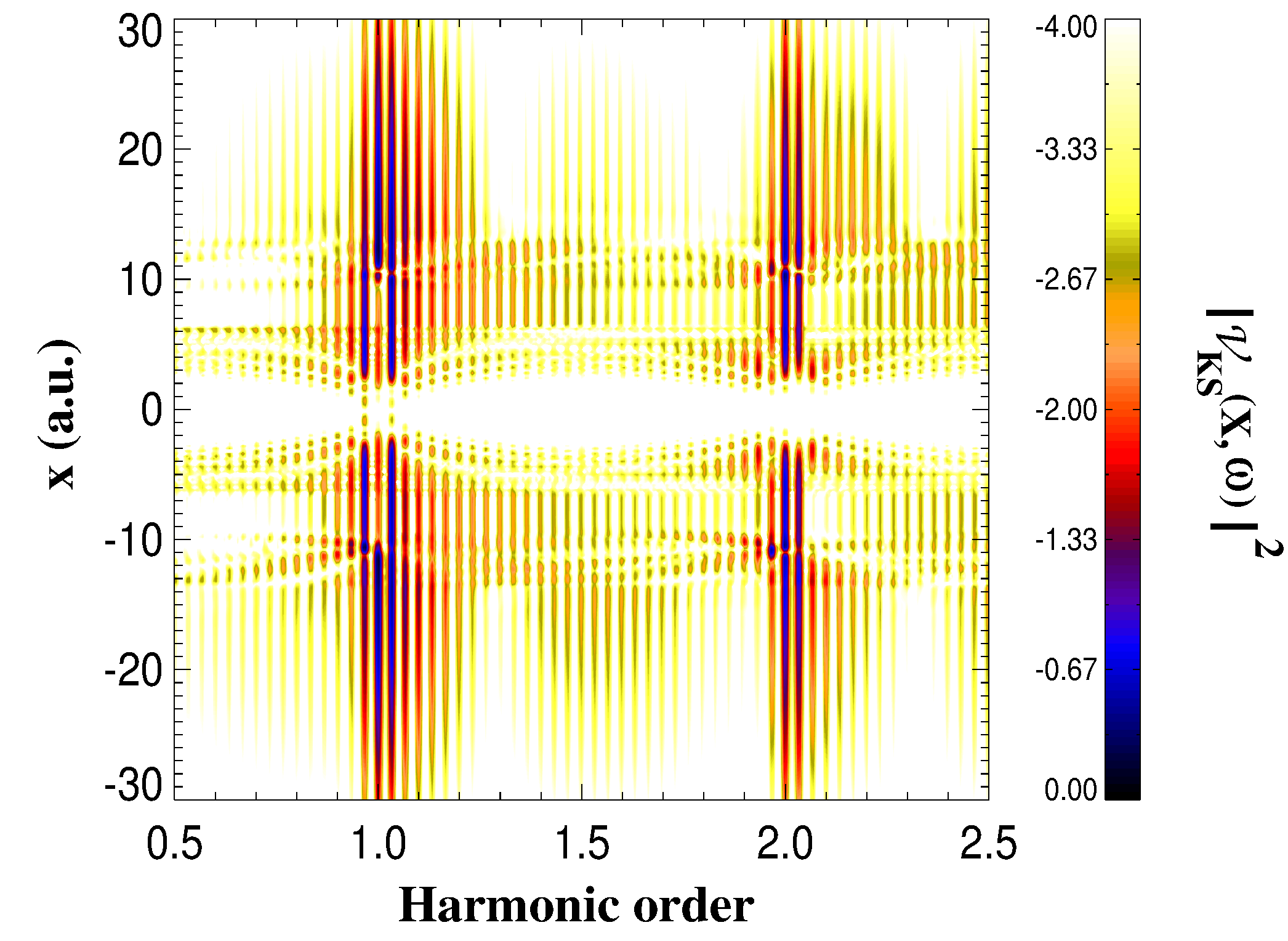} 
\caption{(Color online) As Fig.~\ref{fig:800nm} but for the resonant interaction with $ \omega = \mathcal{E}_{1}-\mathcal{E}_{0}= 0.533$, $\hat{E}=0.016$, two cycles ramp-up, and $148$ cycles constant amplitude.  Peaks at positions $q\omega_{1}\pm p\Omega$ with $q,p \in \mathbb{Z}$ are seen.   \label{fig:rabi}}
\end{figure}

 The laser frequency in our model simulation was tuned to be at resonance between the ground spin-singlet state and the first excited spin-singlet state of the model Helium atom, $ \omega = \mathcal{E}_{1}-\mathcal{E}_{0}= 0.533$ \cite{rabidieter}. For the chosen field amplitude  $\hat{E}=0.016$  (corresponding to a laser intensity of $9 \times 10^{12}$~W/cm$^2$) the ground state population reaches zero at $T_{1/2}\approx 174$, i.e., $\Omega=0.018$. Figure~\ref{fig:rabi} shows $\arrowvert\vKS(x,\omega)\arrowvert^2$ for two cycles ramp-up and $148$ cycles of constant amplitude. The Fourier-transformed potential shows strong sideband peaks at $q\omega_{1}\pm p\Omega$ with $q,p \in \mathbb{Z}$. Hence, while in the previous example of non-adiabatic ramping one might argue that the anharmonic peaks in the spectra are weak and therefore could be ignored, a resonant interaction generates sideband peaks of strengths comparable to the harmonics.

\section{Initial state choice \label{generalresults}}
For the above examples of non-adiabatic ramping or resonant interaction a minimization procedure with a finite Floquet basis would lead to a laser-aperiodic KS Hamiltonian that renders the Floquet theorem inapplicable in the first place. From the TDDFT perspective we obtain  a laser-aperiodic KS Hamiltonian because of the time evolution starting from the chosen initial state. However, in TDDFT $\HopKS([n(t)];t)$ should actually read $\HopKS([n(t),\Psi_0,\{\Phi_{0k}\}];t)$ because of the dependence of the time-dependent KS potential on both the interacting initial state $\Psi_0$ and the KS initial states $\Phi_{0k}$ \cite{TDDFTbooks,inistatedep}. Thus a loophole for a most stubborn assumable proponent of TDDFT-Floquet theory remains: a different choice of initial KS states $\Phi_{0k}(x)=\Phi_k(x,t=0)$ could keep the KS Hamiltonian periodic in $T$. In this Section we give a counter example for which {\em all} possible initial states lead to laser-aperiodic KS potentials if the density is laser-aperiodic. To do so analytically we construct a KS system of two non-interacting electrons on a quantum ring of diameter $L$ so that we have periodic boundary conditions \cite{QR}. This makes it an ideal system to analyze the time-periodicity of the KS Hamiltonian for the various possible  initial states. For spin-singlet states of these electrons one can describe the system by a single KS orbital \reff{KSdensphase} as in our model Helium system above (in the limit $L \rightarrow \infty$ the quantum ring becomes equivalent to the Helium model).  Following the procedure outlined in Sec.~\ref{exactpotential} the potential can be written in terms of the density and the phase of the KS orbital as 
\begin{eqnarray} \lefteqn{\vKS([m,n],x,t)}  \label{potential}\\
&=& \frac{1}{2} \frac{\partial_x^2 \sqrt{n(x,t)}}{\sqrt{n(x,t)}} -\partial_{t}S([m,n],x,t)-\frac{1}{2}[\partial_{x}S([m,n],x,t)]^2,\nonumber \end{eqnarray} 
which is an explicit functional of the density $n=n(x,t)$ and an integer number $m\in\mathbb{Z}$. As shown in Ref.~\cite{QR}, for periodic boundary conditions the phase $S$ can be written in the integral form
\begin{eqnarray} 
S([m,n],x,t)&=&\int_{0}^{L}dy\,K_{t}(x,y)\partial_{t}n(y,t)  \label{phase} \\  &+& \frac{2\pi m}{\int_{0}^{L}\frac{dz}{n(z,t)}} \int_{0}^{x}\frac{dz}{n(z,t)}, \nonumber 
\end{eqnarray} 
where the Green's function, $K_{t}(x,y)$ is defined as
\begin{eqnarray} 
K_{t}(x,y)&=&\frac{1}{2}[\theta(y-x)-\theta(x-y)]\int_{x}^{y}\frac{dz}{n(z,t)} \nonumber \\  &-&\frac{\eta(x,t)\eta(y,t)}{\int_{0}^{L}\frac{dz}{n(z,t)}},\label{kernel}
\end{eqnarray} 
with $\theta$ the Heaviside step function and 
\beq \eta(x,t)=\frac{1}{2}\left(\int_{0}^{x}\frac{dy}{n(y,t)} + \int_{L}^{x}\frac{dy}{n(y,t)} \right). \eeq
Since the KS orbital obeys the periodic boundary conditions, the phase $S$ has to satisfy
\begin{eqnarray}
&S(L,t)&=S(0,t)+2\pi m  \\ &\partial_{x}S(L,t)&=\partial_{x}S(0,t).
\end{eqnarray}
Hence the integer $m$ plays the role of labeling all the possible KS orbitals (for different initial-state choices) that are consistent with the density $n(x,t)$.

If we assume that \beq n(x,t+T)=n(x,t), \eeq we have
\beq \int_{x}^{y}\frac{dz}{n(z,t)}=\int_{x}^{y}\frac{dz}{n(z,t+T)}. 
\eeq
Since $K_{t}(x,y)$ in Eq.~\reff{kernel} consists only of such time-periodic integrals 
\beq
K_{t}(x,y)=K_{t+T}(x,y).
\eeq
Also, since 
\beq
\partial_{t}n(x,t)\arrowvert_{t}=\partial_{t}n(x,t)\arrowvert_{t+T},
\eeq
we conclude from ~\reff{phase} that
\beq
S([m,n],x,t)=S([m,n],x,t+T).
\eeq
The first term on the right hand side of Eq.~\reff{potential} is also periodic with the same period as the density. This implies that the entire potential is periodic with the same period as the density, i.e.,
\beq
\vKS([m,n],x,t)=\vKS([m,n],x,t+T).
\eeq
Hence for any possible initial state (labeled by the index $m$) and a density periodic with the period of the external field we find that the KS Hamiltonian is also periodic with the period of the external field.

For the Floquet theorem to be applicable in a TDDFT framework, the time-dependent KS Hamiltonian must be periodic with the period of the external field $T=2\pi/\omega_1$ only, i.e.,
\beq
\vKS([m,n],x,t)=\vKS([m,n],x,t+T).\label{singlyperiodic}
\eeq 
Consider now the density being periodic with a period  $T'$ different from the period of the external field, 
\beq
n(x,t)=n(x,t+T'), 
\eeq
as in the above examples in Secs.~\ref{nonadiabaticexample} and \ref{resonantinteraction}.
The periods $T$ and $T'$ are incommensurate in general.  We just have proven that the KS potential is periodic with the same period as the density, which implies that
\beq
\vKS([m,n],x,t)=\vKS([m,n],x,t+T^\prime).\label{doublyperiodic}
\eeq 
This is in contradiction with the assumption of only one period $T=2\pi/\omega_1$ of Eq.~\reff{singlyperiodic} which allows the Floquet theorem to be applied in the first place. Hence, the Floquet theorem cannot be applied.

Here, for our example for which we are able to write down an explicit expression for the KS potential, we have proven that for \textit{any} initial KS state it is impossible to have a laser-periodic KS potential when the density has another period. 

\section{Conclusions \label{conclude}}
We investigated the applicability of the Floquet theorem to time-dependent Kohn-Sham  Hamiltonians. 
By employing analytically and numerically exactly solvable counter examples we showed that, in general, Floquet theory is not compatible with time-dependent density functional theory. The reason is that, while periodic drivers such as laser fields of course render the interacting many-body Hamiltonian periodic, the corresponding Kohn-Sham Hamiltonian, in general, is aperiodic. We discussed how the periodicity properties of the single-particle density translate to the Kohn-Sham potential. 
If in the Floquet analysis of the many-body time-dependent Schr\"odinger wave function more than one Floquet state plays a role---such as for non-adiabatic ramping or resonant interactions---the exact Kohn-Sham potential is aperiodic so that the Floquet theorem is inapplicable. Further we showed that also the initial-state dependence of the time-dependent Kohn-Sham Hamiltonian cannot be employed to restore its periodicity. Of course, one may view Kohn-Sham-Floquet theory as an approximative approach for the study of laser-matter phenomena in which resonances and non-adiabaticities are expected to be not relevant.

\section*{Acknowledgment}

This work was supported by the SFB 652 of the German Science Foundation (DFG). M.R. acknowledges  financial support by the Erwin Schr\"odinger Fellowship J 3016-N16 of the FWF (Austrian Science Fund).


\begin{thebibliography}{1}

\bibitem{kohnNobel} W.\ Kohn, Rev.\ Mod.\ Phys.\ {\bf 71}, 1253 (1999).



\bibitem{DFT}
P.\ Hohenberg, W.\ Kohn, Phys.\ Rev.\ \textbf{136}, B864 (1964).

\bibitem{DFTbooks}
See, e.g., R.\ M.\ Dreizler and E.\ K.\ U.\ Gross, \textit{Density Functional Theory, An Approach to the Quantum Many-Body Problem} (Springer, Berlin, 1990).


\bibitem{KS}
W.\ Kohn and L.\ J.\ Sham.\ Phys.\ Rev.\,  \textbf{140}, A1133  (1965).


\bibitem{Runge}
E.\ Runge and E.\ K.\ U. Gross, Phys.\ Rev.\ Lett.\ \textbf{52}, 997 (1984).

\bibitem{TDDFTbooks}
See, e.g.,  C.\ A.\ Ullrich, {\em Time-Dependent Density-Functional Theory} (Oxford University Press, 2012).


\bibitem{Leeuwen}   R.\ van Leeuwen, Phys.\ Rev.\ Lett. \textbf{82}, 3863 (1999). 

\bibitem{ruggibauer} M.\ Ruggenthaler and D.\ Bauer, \pra {\bf 80}, 052502 (2009).



\bibitem{inistatedep} Neepa T.\ Maitra, Kieron Burke, and Chris Woodward, \prl {\bf 89}, 023002 (2002); Neepa T.\ Maitra and Kieron Burke, \pra {\bf 63}, 042501 (2001); \pra {\bf 64}, 039901(E) (2001). 



\bibitem{exampleswhereTDDFTfails} Note that the majority of papers having  TDDFT in the title or abstract actually do not go beyond linear response. 


\bibitem{floquet} M.G.\ Floquet, Ann.\ \'Ecol.\ Norm.\ Sup.\ \textbf{12}, 47 (1883).


\bibitem{floquetclassics} J.H.\ Shirley, Phys.\ Rev.\ \textbf{138}, B979 (1965); H.\ Sambe, \pra  {\bf 7},   2203 (1973).

\bibitem{floquetinbooks} Floquet theory is covered in several text books, e.g., D.J.\ Tannor, {\em Introduction to Quantum Mechanics: a Time-Dependent Perspective} (University Science Books, Sausalito, 2007); B.H.\ Bransden, C.J.\ Joachain, {\em Physics of Atoms and Molecules} (Prentice Hall, Harlow, 2003); C.J.\ Joachain, N.J.\ Kylstra, R.M.\ Potvliege, {\em Atoms in Intense Laser Fields} (Cambridge University Press, Cambridge, 2012); 
H.\ Friedrich, {\em Theoretical Atomic Physics}, (Springer, Berlin, 2006); F.H.M.\ Faisal, {\em Theory of Multiphoton Processes} (Plenum Press, New York, 1987). Bloch states in space-periodic potentials may be viewed as ``Floquet states in space.''


\bibitem{floquetanalysis}
V.\ Kapoor and D.\ Bauer,   Phys.\ Rev.\ A \textbf{85}, 023407 (2012).


\bibitem{deb}

B.\ M.\ Deb, S.\ K.\ Ghosh, J. Chem.\ Phys.\ \textbf{77}, 342 (1982).

\bibitem{Hone}
D.\ W.\ Hone, R.\ Ketzmerick, and W.\ Kohn, Phys.\ Rev.\ A \textbf{56}, 4045 (1997). 


\bibitem{TDDFTstates1}
D.A.\ Telnov, S.- I.\ Chu. Chem.\ Phys. Lett.\ \textbf{264}, 466 (1997).



\bibitem{examplerev}
Pawel Salek, Trygve Helgaker, Trond Saue, Chem.\ Phys. Lett.\ \textbf{311},187 (2005).

\bibitem{floquetreview} 
S.-I.\ Chu, D.A.\ Telnov, Phys.\ Rep.\ {\bf 390}, 1 (2004).
\bibitem{TDDFTstates2}

D.A.\ Telnov, S.-I.\ Chu. Phys.\ Rev.\ \textbf{58}, 6 (1998).

\bibitem{floquetcritic}
N.\ T.\ Maitra, K.\ Burke, Chem.\ Phys. Lett.\ \textbf{359},  237 (2002).


\bibitem{criticfinite}
N.T.\ Maitra, K.\ Burke, Chem.\ Phys. Lett.\ \textbf{441}, 167 (2007).





\bibitem{helium}
See, e.g., R. Grobe and J. H. Eberly, Phys.\ Rev.\ Lett. \textbf{68}, 2905 (1992); S. L. Haan, R. Grobe, and J. H. Eberly, Phys.\ Rev.\ A \textbf{50}, 378 (1994); D. Bauer, Phys.\ Rev.\ A \textbf{56}, 3028
(1997); D. G. Lappas and R. van Leeuwen, J.\ Phys.\ B \textbf{31},
L249 (1998); M. Lein, E. K. U. Gross, and V. Engel, Phys.\
Rev.\ Lett. \textbf{85}, 4707 (2000).

\bibitem{exactpotential2}
M.\ Lein and S.\ K\"ummel, Phys.\ Rev. Lett.\ \textbf{94}, 143003 (2005).

\bibitem{rabidieter}
M.\ Ruggenthaler and D.\ Bauer, Phys.\ Rev.\ Lett.\ \textbf{102}, 233001 (2009). 



\bibitem{exactpotential1}
I.\ D’Amico and G.\ Vignale, Phys.\ Rev.\ B \textbf{59}, 7876 (1999).

\bibitem{QR}
M.\ Ruggenthaler, S.\ E.\ B.\ Nielsen, R.\ van Leeuwen, arXiv:1209.2949v2.







\bibitem{Godby}
J.\ D.\ Ramsden and R.\ W.\ Godby,  Phys.\ Rev. Lett.\ \textbf{109}, 036402 (2012).


\bibitem{Soeren}
S.\ E.\ B.\ Nielsen, M.\ Ruggenthaler, R.\ van Leeuwen, Europhys.\ Lett {\bf 101}, 33001 (2013). 




\bibitem{helbig}
J.\ I.\ Fuks, N.\ Helbig, I.\ V.\ Tokatly, and A.\ Rubio,  Phys.\ Rev.\ B \textbf{84}, 075107 (2011).







\end{thebibliography}
\end{document}